\documentclass{article}
\usepackage{graphicx}
\newcommand{\alt}{\mathrel{\raisebox{-.6ex}{$\stackrel{\textstyle<}{\sim}$}}}

\def\err#1#2{$\stackrel{\scriptstyle +#1}{\scriptstyle -#2}$}

\begin{document}
\centerline{Date:\today     \hfill{\bf UCD-HEP-99-10} }
\vspace*{.5in}
\begin{center}
{\large Global Lepton-Quark Neutral Current Constraint on Low Scale Gravity
Model} \\
\vskip 0.7cm
Kingman Cheung \\
{\it Department of Physics, University of California, Davis, 
CA 95616 USA} \\
\date{\today}
\end{center}

\begin{abstract}
We perform a global analysis of the lepton-quark neutral current data on the
low scale gravity model, which arises from the extra dimensions.  The global
data include HERA neutral current deep-inelastic scattering, 
Drell-yan production at the Tevatron, and fermion pair production at LEPII.
The Drell-yan production, due to the large invariant mass data, provides
the strongest constraint.  Combining all data, the effective Planck scale
must be larger than about 1.12 TeV for $n=3$ and 0.94 TeV for $n=4$ at 
95\%CL. 
\end{abstract}

\section{Introduction}

A solution to the gauge hierarchy, based on extra spacial 
dimensions,  was recently proposed by Arkani-Hamed, Dimopoulos and Dvali
\cite{theory}.  They assumed the space-time is $4+n$ dimensional, with the
standard model (SM) particles living on a brane.  While the electromagnetic, 
strong, and weak forces are confined to this brane, gravity can propagate 
in the extra dimensions.  To solve the gauge hierarchy problem they
proposed the ``new'' Planck scale $M_S$ is of the order of 
TeV with very large extra dimensions.  The size $R$ of these extra dimensions
can be as large as 1 mm, which corresponds to a compactification scale 
$R^{-1}$ as low as $10^{-4}$ eV.   The usual Planck scale 
$M_G=1/\sqrt{G_N} \sim 1.22 \times 10^{19}$ GeV is related to this effective
Planck scale $M_S$ using the Gauss's law:
\begin{equation}
R^n \, M_S^{n+2} \sim M_G^2 \;,
\end{equation}
where $n$ is the number of extra dimensions.
For $n=1$ it gives a large value for $R$, which is already ruled out 
by gravitational
experiments.  On the other hand, $n=2$ gives $R \alt 1$ mm, which is 
in the margin beyond the reach of present gravitational experiments.

The graviton including its excitations in the extra dimensions 
can couple to the SM particles on the brane with an effective strength of
$1/M_S$ (instead of $1/M_G$) after summing the effect 
of all excitations collectively, and thus the gravitation interaction
becomes comparable in strength to weak interaction at TeV scale.  Hence, it
can give rise to a number of phenomenological activities testable at existing
and future colliders \cite{collider}.  
So far, studies show that there
are two categories of signals: direct and indirect.  The indirect signal
refers to exchanges of gravitons in the intermediate states, while direct
refers to production or associated production of gravitons in the final 
state.   Indirect signals include fermion pair,
gauge boson pair production, correction to precision variables, etc. 
There are also other astrophysical and cosmological signatures and 
constraints \cite{others}.  Among the constraints the cooling of supernovae
by radiating gravitons places the strongest limit on the effective Planck
scale $M_S$ of order 50 TeV for $n=2$, which renders collider signatures 
for $n=2$ uninteresting.  Thus, we concentrate on $n\ge 3$. 

In this work, we perform a global analysis of the lepton-quark neutral 
current data on the low scale gravity model. The global
data include HERA neutral current deep-inelastic scattering, 
Drell-yan production at
the Tevatron, and total hadronic, $b\bar b$ and $c\bar c$ pair cross sections
at LEPII.  In addition, we also include the leptonic cross sections 
$e^+ e^- \to \mu^+\mu^-, \tau^+ \tau^-$ at LEPII in our analysis.
The $\nu$-N scattering data from CCFR and NuTeV have been shown
by Rizzo \cite{collider} to be very insignificant in constraining the low 
scale gravity, and so we shall not include this data set in our analysis.
We shall see that 
the Drell-yan production, due to the large invariant mass data, provides
the strongest constraint among the global data.  
By combining all data, the effective Planck scale $M_S$ must be larger 
than about 1.12 TeV for $n=3$ and 0.94 TeV for $n=4$ at 95\%CL.  
This is our main result.

The organization of the paper is as follows.  In the next section, we 
shall describe each set of data used in our global analysis and derive the
effect of the low scale gravity.
In Sec. III, we give the numerical results and interpretations for our fits, 
{}from which we can draw the limits on the effective Planck scale.

\section{Global Data}

Before we come to each data set, let us first give a general expression
for the square of amplitude for $e^- e^+ \to q \bar q$:
\begin{eqnarray}
\sum \left| {\cal M} \right |^2 &=&
4 u^2 \left( |M_{LL}|^2 + |M_{RR}|^2 \right) +
4 t^2 \left( |M_{LR}|^2 + |M_{RL}|^2 \right)  \nonumber \\
&+& 2\pi^2 \; \left( \frac{\cal F}{M_S^4} \right )^2 \;
    ( t^4 - 6 t^3 u + 18 t^2 u^2 - 6 t u^3 + u^4 ) \nonumber \\
&+& 8 \pi e^2 Q_e Q_q \;  \frac{\cal F}{M_S^4} \; \frac{(u-t)^3}{s} 
  \nonumber \\
&+& \frac{8 \pi e^2}{\sin^2\theta_{\rm w} \cos^2 \theta_{\rm w} } \;
 \frac{\cal F}{M_S^4} \; \frac{1}{s - M_Z^2} \;
\biggr[  g_a^e g_a^q ( t^3 - 3t^2 u - 3t u^2 + u^3 )
       + g_v^e g_v^q ( u-t)^3 \biggr ] \;,
\end{eqnarray}
where $s,t,u$ are the usual Mandelstam variables, $e = \sqrt{4\pi \alpha}$ 
is the electromagnetic coupling constant, $Q_f$ is the 
electric charge of the fermion $f$, $\theta_{\rm w}$ is the Weinberg mixing
angle, $g_a^f$ and $g_v^f$ are, respectively, the axial-vector and the
vector $Z$ couplings to the fermion $f$.  
The reduced amplitudes $M_{\alpha\beta}$ ($\alpha,\beta= L,R$) are
\begin{equation}
\label{reduce}
M_{\alpha\beta} = \frac{e^2 Q_e Q_q}{s} + 
\frac{e^2}{\sin^2\theta_{\rm w} \cos^2 \theta_{\rm w} } \;
\frac{g_\alpha^e g_\beta^q}{s - M_Z^2}
\end{equation}
where 
\begin{eqnarray}
g_L^f = T_{3f} - Q_f \sin^2\theta_{\rm w}\;, &&
g_R^f = - Q_f \sin^2\theta_{\rm w} \nonumber \\
g_v^f = (g_L^f + g_R^f)/2\;, && g_a^f = (g_L^f - g_R^f)/2 \nonumber \;.
\end{eqnarray}
In the derivation, we have followed the notation in Han {\it et al}. 
\cite{collider} and 
we have taken $M_S^2 \gg s, |t|, |u|$ and in this case the propagator
factor $D(s)=D(|t|)=D(|u|)$, which is given by
\begin{equation}
\kappa^2 |D(s)| = \frac{16 \pi}{M_S^4}\; \times {\cal F} \;,
\end{equation}
where the factor ${\cal F}$ is given by
\begin{equation}
\label{F}
{\cal F} = \Biggr \{ 
\begin{array}{l}
\log \left( \frac{M_S^2}{s} \right ) \;\; {\rm for}\;\; n=2 \;, \\
\frac{2}{n-2} \;\;\;\;\;\;\;\;\;\;\; {\rm for}\;\; n>2 \;.
\end{array}
\end{equation}
For $n\ge 3$ ${\cal F}$ is always positive.
The amplitude square for the crossed channels $e^\pm \stackrel{(-)}{q} \to
e^\pm \stackrel{(-)}{q}$ can be obtained using the crossing symmetry.

\subsection{HERA neutral-current data}

Both H1 and ZEUS have measured the neutral-current deep-inelastic
scattering cross sections at high-$Q^2$ region.
Despite the excess in cross section reported in early 1997, the 1997 data 
alone agreed well with the SM.  The double differential cross section 
for $e^+ p \to e^+ X$ is given by
\begin{eqnarray}
\frac{d^2\sigma}{dy dQ^2} &=& \sum_q
\frac{f_q(x)}{16\pi}\, \frac{Q^2}{s y^2} \; \Biggr \{
(1-y)^2 ( |M_{LL}|^2 + |M_{RR}|^2)  + |M_{LR}|^2 + |M_{RL}|^2 \nonumber  \\
&&+
\frac{\pi^2}{2}\, \left( \frac{Q^2}{y} \right )^2 \, 
   \left( \frac{\cal F}{M_S^4} \right )^2 \; 
   (32 - 64 y +42 y^2 -10y^3 + y^4 ) \nonumber \\
&&+ 
2 \pi e^2 Q_e Q_q \;\left( \frac{\cal F}{M_S^4} \right )\;
  \frac{(2-y)^3}{y} \nonumber \\
&&+
\frac{2\pi e^2 }{\sin^2\theta_{\rm w} \cos^2\theta_{\rm w}}\;
  \left( \frac{\cal F}{M_S^4} \right )\;
  \left( \frac{Q^2}{y} \right )\; \frac{1}{-Q^2  - M_Z^2} \; \biggr[
g_a^e g_a^q (6y -6y^2 +y^3 ) + g_v^e g_v^q (y-2)^3  \biggr ] \Biggr \} 
\nonumber \\
&&+
\frac{\pi}{2}\, f_{g}(x) \, \frac{Q^2}{s y^2} \; 
   \left( \frac{\cal F}{M_S^4} \right )^2 \; 
\left( \frac{Q^2}{y} \right)^2  \; (1-y)(y^2 -2y + 2) \;,
\end{eqnarray}
where $Q^2 =s x y$ is the square of the momentum-transfer, $\sqrt{s}=300$ GeV,
and $f_{q/g}(x)$ are the parton distribution functions.
The reduced amplitudes $M_{\alpha\beta}$ are given in Eq. (\ref{reduce})
with $s$ replaced by $-Q^2$. 
The last term in the above equation 
comes from the additional channel $e^+ g \to e^+ g$ allowed by
the low scale gravity interactions.  The channels
$\sum_{\bar q} e^+ \bar q \to e^+ \bar q$ are also included.
The $Q^2$ data of ZEUS \cite{h1-zeus}, including systematics, are 
recently published
while the H1 data \cite{h1-zeus} used in our analysis are shown in 
Table \ref{hera}.

\subsection{Drell-yan at the Tevatron}

CDF measured the double differential cross section $d^2\sigma/dM_{\ell\ell}
dy\;(\ell=e,\mu)$ 
for the Drell-yan production, where $M_{\ell\ell}$ and $y$ are,
respectively, the invariant mass and the rapidity of the lepton pair.  
Essentially, CDF measured the cross section in the range $-1<y<1$ and then
average over $y$ to obtain the double differential cross section. 
The double differential cross section for $p \bar p \to \ell^+ \ell^-$ 
is given by
\begin{eqnarray}
\frac{d^2\sigma}{dM_{\ell\ell} dy} &=& K \Biggr\{ \sum_q \; \biggr [
\frac{M_{\ell\ell}^3}{72\pi s} \;
f_q(x_1) f_{\bar q}(x_2)\; \left(
|M_{LL}|^2 + |M_{LR}|^2 + |M_{RL}|^2 + |M_{RR}|^2  \right ) \nonumber \\
&+&
\frac{\pi M_{\ell\ell}^7}{120 s}\; f_q(x_1) f_{\bar q}(x_2) \;
\left( \frac{\cal F}{M_S^4} \right )^2  \; \biggr ]
+
\frac{\pi M_{\ell\ell}^7}{80 s}\; f_g(x_1) f_g (x_2) \;
\left( \frac{\cal F}{M_S^4} \right )^2  \Biggr \} \;,
\end{eqnarray}
where $x_{1,2}=M_{\ell\ell} e^{\pm y}/\sqrt{s}$, $\sqrt{s}=1800$ GeV, 
$f_{q,g}(x_i)$ are the parton distribution functions,
and the sum over all possible $q\bar q$ pairs is
understood.  
Note that after integrating the scattering angle $\cos\theta^*$ over the 
whole range in the center-of-mass frame, the interference terms 
proportional to $({\cal F}/M_S^4)$ vanish. 
The reduced amplitudes $M_{\alpha\beta}\;(\alpha,\beta=L,R)$
are given in Eq. (\ref{reduce}) with $s$ replaced by $\hat s=M_{\ell\ell}^2$.
The QCD $K$ factor is given by $K=1+ \frac{\alpha_s(\hat s)}{2\pi}\frac{4}{3}
( 1+ \frac{4\pi^2}{3})$.
The second term of the above equation comes from 
$q\bar q \to G \to \ell^+ \ell^-$ and the third one comes from
$gg \to G \to \ell^+ \ell^-$.  
This $K$ factor is, in principle, not valid for the $gg$ channel, but
it will not affect our calculation because we normalize our SM calculation 
to the expected number of events in each bin given by CDF when we deal with
the CDF data (similarly, we normalize to the first bin of the D0 data
when we deal with D0 data.)
On the other hand, D0 measured the differential cross section $d\sigma/dM_{ee}$
with the rapidity range integrated. The Drell-yan data \cite{dy-cdf-d0} 
we used in our analysis are shown in Table \ref{dy}.

\subsection{LEPII fermion pair Cross sections}

LEPII has measured the leptonic cross sections, hadronic cross sections,
and heavy flavor ($b$ and $c$) production.  Since the low scale gravity 
interactions are naturally flavor-blind, we shall include the effects on all 
flavors here.  For a massless $q$ the expression for 
$\sigma(e^+ e^- \to q \bar q)$ is given by
\begin{equation}
\sigma(e^+ e^- \to q \bar q ) = K \Biggr\{
\frac{s}{16\pi} \left( 
|M_{LL}|^2 + |M_{LR}|^2 + |M_{RL}|^2 + |M_{RR}|^2  \right ) 
+ \frac{3\pi s^3}{80} \left( \frac{\cal F}{M_S^4} \right )^2 \Biggr \} \;,
\end{equation}
where the reduced amplitudes $M_{\alpha\beta}\;(\alpha,\beta=L,R)$
are given in Eq. (\ref{reduce}), and the QCD $K$ factor is 
$K=1+ \alpha_s/\pi + 1.409 (\alpha_s/\pi)^2 -12.77 (\alpha_s/\pi)^3$.
The leptonic cross sections $\sigma(e^+ e^- \to \mu^+\mu^-, \tau^+ \tau^-)$
do not have this $K$ factor.
The LEPII data \cite{lep} 
used in our analysis are given in Tables \ref{lep-h} and \ref{lep-l}.

\subsection{$\nu$-N Scattering}

The recent measurements by CCFR and NuTeV \cite{ccfr} are the most precise
ones on the neutrino-quark couplings.  Since the gravitons also couple to
neutrinos, the measurements should, in principle, place a constraint on
the scale $M_S$.  However, the analysis by Rizzo in Ref. \cite{collider} 
showed that
the constraint coming from these low energy $\nu$-N scattering is very weak.
This is because the center-of-mass energies of these collisions are only
of order of tens of GeV's, even though the neutrino beam energy is as high as 
a few hundred GeV.  In addition, the effective operators induced by the 
low scale 
gravity interactions are at least of dimension eight.  This is in contrast 
to the traditional four-fermion contact interactions, which are only of
dimension six.  Therefore, at such low center-of-mass energies in these
collisions, the effect of low scale gravity is extremely minimal.
We are not going to include these data in our global fit.

\section{Fits}

For the fitting we follow the procedures in Refs. \cite{contact}, where
the four-fermion contact interactions are analyzed with a similar but larger
global data set.  The difference is that Refs. \cite{contact} include also
the data sets from the 
low energy $e$-N scattering and the atomic parity violation, which constrain
parity violating interactions.  But the gravity is certainly parity 
conserving and thus it receives no restriction from these parity-violating
experiments.

The interference effect between the SM amplitude and the low scale gravity
scales as ${\cal F}/M_S^4$, while the pure low scale gravity scales as
$({\cal F}/M_S^4)^2$.  To linearize the fitting we use the parameter 
$\eta={\cal F}/M_S^4$.  We use MINUIT to minimize the $\chi^2$.  The best
estimates of $\eta$ for each individual data set and the combined set are
shown in Table \ref{fit}.  
The SM fit gives a $\chi^2=94.10$ for 118 d.o.f., while the fit with non-zero
$\eta$ gives a $\chi^2=94.05$ for 117 d.o.f.  
Since the fits do not show any evidence for new physics, we 
can then place limits on the $\eta={\cal F}/M_S^4$.  
The physical region of $\eta$ is $\eta \ge 0$, and so we define the 95\%CL
upper limit for $\eta$ as
\begin{equation}
0.95 = \frac{\int_0^{\eta_+} {\cal P}(\eta) \; d\eta}
{\int_0^\infty {\cal P}(\eta) \; d\eta}\;,
\end{equation}
where
\begin{equation}
{\cal P}(\eta) = \frac{1}{\sigma \sqrt{2\pi}}\; \exp \left( -\,
\frac{\chi^2(\eta) - \chi^2_{\rm min} }{2} \right ) \;.
\end{equation}
{}From $\eta_+$ we find the limits as $M_S/{\cal F}^{1/4} = \eta_+^{-1/4}$.  
The limits on $M_S/{\cal F}^{1/4}$ for each data set and the combined 
are also shown in Table \ref{fit}.  The combined limit is 
$M_S/{\cal F}^{1/4} > 939$ GeV, or 
\begin{equation}
\label{result}
M_S > \Biggr\{ \begin{array}{cc}
          1120 \; {\rm GeV}\;\; &{\rm for}\;\; n=3 \\
           939 \; {\rm GeV}\;\; & {\rm for}\;\; n=4  \\
               \end{array}
\end{equation}
at 95\%CL.

\begin{figure}[th]
\leavevmode
\begin{center}
\includegraphics[height=4.5in]{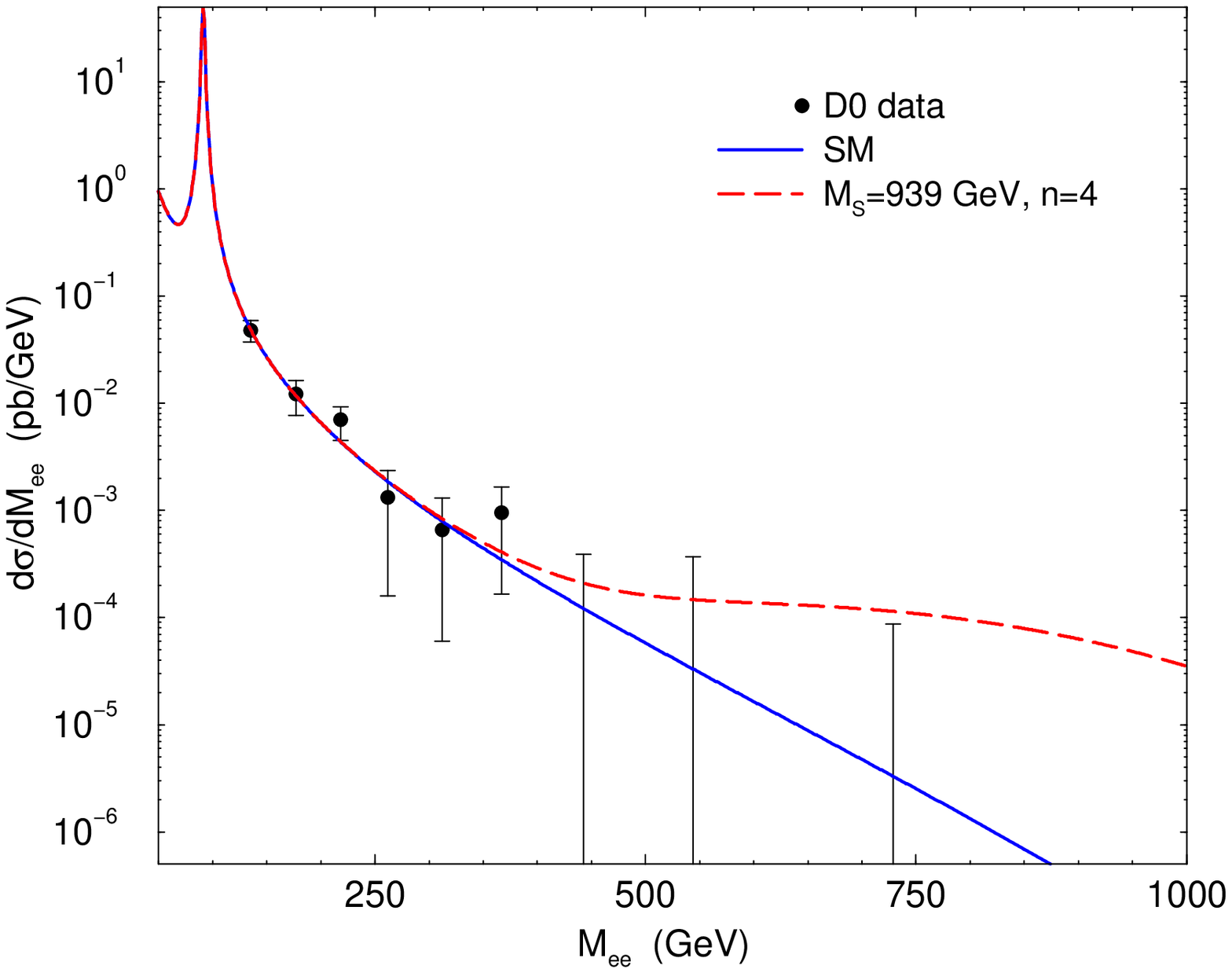}
\end{center}
\caption{
The differential distribution $d\sigma/d M_{ee}$ for
Drell-yan production at the Tevatron for the low scale gravity model and 
for the SM.  The data shown are from D0.
}
\label{fig1}
\end{figure}

It is the Drell-yan production at the Tevatron that 
gives the strongest constraint among all data.  This is easy to
understand because the $\hat s$ of the Drell-yan process is the largest 
of all.  The new HERA data are now consistent with the SM 
and only cause a very slight pull of the fit, about one third of a 
$\sigma$ from zero: see Table \ref{fit}.
Both the ``ALL w/o HERA'' and the ``ALL'' fits agree well with the SM.
We show in Fig. \ref{fig1} the Drell-yan production for the SM and for
the low scale gravity model with $M_S$ and $n$ given in Eq. (\ref{result}), 
together with the D0 data.

To conclude, we have used the global lepton-quark neutral current data plus
the leptonic cross sections at LEPII to constrain the low scale gravity
interactions arising from the extra dimensions.  The limit that we place
on the effective Planck scale $M_S$ is about 1.12 TeV for $n=3$ and 0.94 TeV 
for $n=4$ at 95\%CL.  

\section*{\bf Acknowledgments}
This research was supported in part by the U.S.~Department of Energy under
Grants No. DE-FG03-91ER40674 and by the Davis Institute for High Energy 
Physics.  


\begin{table}[th]
\caption{ \label{hera}
\small The measured $N_{\rm obs}$ and expected $N_{\rm exp}$ 
number of events as a function of $Q^2_{\rm min}$ for H1 at HERA.}
\medskip
\centering
\begin{tabular}{|ccl|}
\hline
\multicolumn{3}{|c|}{H1 (${\cal L}=37.04\, {\rm pb}^{-1}$) } \\
\hline
\hline
\underline{$Q^2_{\rm min} \,({\rm GeV}^2)$} & \underline{$N_{\rm obs}$} &
\underline{$N_{\rm exp}$} \\
  2500 & 1297 & 1276$\pm98$ \\ 
  5000  & 322  & 336$\pm29.6$ \\
 10000  & 51   & 55.0$\pm6.42$ \\
 15000  & 22   & 14.8$\pm2.13$ \\
 20000 & 10   & 4.39$\pm0.73$ \\
 25000 & 6    & 1.58$\pm0.29$ \\
\hline
\end{tabular}
\end{table}

\begin{table}[th]
\caption{ \label{dy}
\small The Drell-Yan data by CDF and D0 \cite{dy-cdf-d0}. $N_{\rm obs}$ is 
the observed and $N_{\rm exp}$ is the expected number of events.}
\medskip
\centering
\begin{tabular}{|c@{\extracolsep{0.2in}}||cc|cc|}
\hline
\multicolumn{5}{|c|}{CDF} \\
\hline
\hline
& 
\multicolumn{2}{c|}{$e^+ e^-$} & \multicolumn{2}{c|}{$\mu^+ \mu^-$} \\
\hline
\underline{$M_{\ell\ell}$ (GeV)} & \underline{$N_{\rm obs}$}
 & \underline{$N_{\rm exp}$} & \underline{$N_{\rm obs}$}
 & \underline{$N_{\rm exp}$}  \\
50--150  &  2581 & 2581 & 2533 & 2533 \\ 
150--200 &  8    & 10.8 &   9  & 9.7  \\
200--250 &  5    & 3.5  &   4  & 3.2  \\
250--300 &  2    & 1.4  &   2  & 1.3  \\
300--400 &  1    & 0.97 &   1  &  0.94 \\
400--500 &  1    & 0.25 &   0  &  0.27 \\
500--600 &  0    & 0.069 &  0  & 0.087 \\
\hline
\end{tabular}

\vspace{1in}

\begin{tabular}{|c@{\extracolsep{0.2in}}||c|}
\hline
\multicolumn{2}{|c|}{D0} \\
\hline
\hline
   &  $e^+ e^-$ \\
\hline
\underline{$M_{\ell\ell}$ (GeV)} & \underline{$\sigma$ (pb)}  \\
120--160 & 1.93 \err{0.43}{0.44} \\ 
160--200 & 0.49 \err{0.16}{0.18} \\ 
200--240 & 0.28 \err{0.09}{0.10} \\ 
240--290 & 0.066 \err{0.052}{0.058} \\  
290--340 & 0.033 \err{0.032}{0.030} \\ 
340--400 & 0.057 \err{0.042}{0.047} \\ 
400--500 & $<0.063 \;(0.039)$ \\
500--600 & $<0.060 \;(0.037)$ \\
600--1000 & $<0.058 \;(0.035)$ \\
\hline
\end{tabular}
\end{table}

\begin{table}[tbh]
\caption{ \label{lep-h}
\small 
The LEPII  hadronic cross sections and the ratio $R_b$ and $R_c$. Note that
the cross sections given by ALEPH have an angular cut of $|\cos\theta|<0.95$.
The two sets of DELPHI data for $R_b,R_c$ are given at 
$\sqrt{s}=130-136$ and $161-172$ GeV, respectively.
$\sigma^{\rm SM}_{\rm had}$ and $R_{b,c}^{\rm SM}$ are the SM expectations.}
\medskip
\centering
\begin{tabular}{|c@{\extracolsep{0.1in}}||cc|cc|cc|}
\hline
$\sqrt{s}$ (GeV) &$\sigma_{\rm had}$ (pb)& 
           $\sigma^{\rm SM}_{\rm had}$ (pb) &
$R_b$ & $R_b^{\rm SM}$ & $R_c$ & $R_c^{\rm SM}$ \\
\hline
\hline
\multicolumn{7}{|c|}{ALEPH} \\
\hline
130 & 71.6 $\pm$ 3.96 &  70.7 & 0.176 $\pm$ 0.044 & 0.190 & - & - \\
136 & 58.8 $\pm$ 3.61 &  57.3 & 0.214 $\pm$ 0.050 & 0.186 & - & - \\
161 & 29.94$\pm$ 1.84 &  30.7 & 0.159 $\pm$ 0.035 & 0.175 & - & - \\
172 & 26.4 $\pm$ 1.75 &  25.1 & 0.134 $\pm$ 0.037 & 0.173 & - & - \\
183 & 21.71$\pm$ 0.74 &  21.1 & 0.176 $\pm$ 0.020 & 0.171 & - & - \\
189 & 19.59 $\pm$ 0.46 & 19.3 &  -                & -     & - & - \\
\hline
\multicolumn{7}{|c|}{DELPHI}\\
\hline
130.2 & 82.1 $\pm$ 5.2 &  83.1 & 0.174 $\pm$ 0.028 & 0.182 & 0.199$\pm$0.073 
 & 0.225 \\
136.2 & 65.1 $\pm$ 4.7 &  67.0 &                   &       & &\\
161.3 & 40.9 $\pm$ 2.1 &  34.8 & 0.142 $\pm$ 0.024 & 0.165 & 0.314$\pm$0.055 &
 0.250 \\ 
172.1 & 30.3 $\pm$ 1.9 &  28.9 &                  &      &   & \\
\hline
\multicolumn{7}{|c|}{L3}\\
\hline
130.3 & 81.8 $\pm$ 6.4 &  78 &  -                & -     & - & - \\
136.3 & 70.5 $\pm$ 6.2 &  63 &  -                & -     & - & - \\
140.2 & 67   $\pm$ 47  &  56 &  -                & -     & - & - \\
161.3 & 37.3 $\pm$ 2.2 &  34.9 &  -                & -     & - & - \\ 
170.3 & 39.5 $\pm$ 7.5 &  29.8 &  -                & -     & - & - \\
172.3 & 28.2 $\pm$ 2.2 &  28.9 &  -                & -     & - & - \\
\hline
\multicolumn{7}{|c|}{OPAL} \\
\hline
130.25 & 64.3 $\pm$ 5.1 &  77.6 &  -                & -     & - & - \\
133.17 & -              &   -   & 0.195 $\pm$ 0.041 & 0.182 & - & - \\
136.22 & 63.8 $\pm$ 5.2 &  62.9 &  -                & -     & - & - \\
161.34 & 35.5 $\pm$ 2.2 &  33.7 & 0.162 $\pm$ 0.041 & 0.169 & - & - \\ 
172.12 & 27.0 $\pm$ 1.9 &  27.6 & 0.131 $\pm$ 0.047 & 0.165 & - & - \\
183    & 23.7 $\pm$ 0.81&  24.3 & 0.190 $\pm$ 0.022 & 0.163 & - & - \\
189    & 22.1 $\pm$ 0.64&  22.3 & 0.167 $\pm$ 0.014 & 0.162 & - & - \\
\hline
\end{tabular}
\end{table}

\begin{table}[tbh]
\caption{\label{lep-l}
\small
The LEPII leptonic cross sections for $e^+ e^- \to \mu^+ \mu^-, \tau^+ \tau^-$.
Note that the cross sections given by ALEPH have an angular cut of 
$|\cos\theta|<0.95$.
$\sigma^{\rm SM}$ are the SM expectations.}
\medskip
\centering
\begin{tabular}{|c@{\extracolsep{0.1in}}||cc|cc|}
\hline
$\sqrt{s}$ (GeV) & 
$\sigma_{\rm \mu\mu}$ (pb)& $\sigma^{\rm SM}_{\rm \mu\mu}$ (pb) &
$\sigma_{\rm \tau\tau}$ (pb)&$\sigma^{\rm SM}_{\rm \tau\tau}$ (pb)\\
\hline
\hline
\multicolumn{5}{|c|}{ALEPH} \\
\hline
130 &  7.9 $\pm$ 1.2  &  7.0  &  10.9 $\pm$ 1.8 & 7.3   \\
136 &  6.9 $\pm$ 1.1  &  6.1  &   5.6 $\pm$ 1.3 & 6.3   \\
161 &  4.49$\pm$ 0.70 &  3.88 &  5.75 $\pm$ 0.99 & 4.01  \\
172 &  2.64$\pm$ 0.53 &  3.32 &  3.26 $\pm$ 0.75 & 3.43  \\
183 &  2.98$\pm$ 0.25 &  2.89 &  2.90 $\pm$ 0.30 & 2.98  \\
189 &  2.66$\pm$ 0.14 &  2.69 &  2.39 $\pm$ 0.16 & 2.78  \\
\hline
\multicolumn{5}{|c|}{DELPHI}\\
\hline
130.2 & 9.7 $\pm$ 1.9 &  8.1 & 10.2 $\pm$ 3.1 & 8.3  \\
136.2 & 6.6 $\pm$ 1.6 &  7.0 & 8.8  $\pm$ 3.0 & 7.2  \\
161.3 & 3.6 $\pm$ 0.7 &  4.5 & 5.1 $\pm$ 1.2 &  4.6  \\ 
172.1 & 3.6 $\pm$ 0.7 &  3.8 & 4.5 $\pm$ 1.1 & 3.9   \\
\hline
\multicolumn{5}{|c|}{L3}\\
\hline
161.3 & 4.59 $\pm$ 0.84 &  4.4 &  4.6 $\pm$ 1.1   & 4.5 \\ 
172.1 & 3.60 $\pm$ 0.75 &  3.8 &  4.3 $\pm$ 1.1   & 3.9 \\
\hline
\multicolumn{5}{|c|}{OPAL} \\
\hline
130.25 & 9.0  $\pm$ 1.8 &  8.0 &  6.8 $\pm$ 2.0  & 8.0 \\
136.22 & 11.4 $\pm$ 2.1 &  7.0 &  7.2 $\pm$ 2.1  & 6.9 \\
161.34 & 4.5  $\pm$ 0.7 &  4.4 &  6.2 $\pm$ 1.0  & 4.4 \\ 
172.12 & 3.6  $\pm$ 0.6 &  3.8 &  3.9  $\pm$ 0.8 & 3.8 \\
183    & 3.46 $\pm$ 0.29&  3.45 & 3.31 $\pm$ 0.32 & 3.45 \\
189    & 3.13 $\pm$ 0.16&  3.21 & 3.53 $\pm$ 0.21 & 3.21  \\
\hline
\end{tabular}
\end{table}

\begin{table}[tbh]
\caption{ \label{fit}
\small
The best estimate of the $\eta={\cal F}/M_S^4$, 95\%CL upper limit
$\eta_+$, and the 95\%CL lower limit on $M_S/{\cal F}^{1/4}$.}
\medskip
\centering
\begin{tabular}{|c|cc|c|}
\hline
Data Set  &  $\eta={\cal F}/M_S^4$ (GeV$^{-4}$)  &  $\eta_+$ (GeV$^{-4}$)  & 
95\%CL $M_S/{\cal F}^{1/4}$\\
\hline
\hline
DY-CDF  &  $(0.00 \pm 0.15)\times 10^{-11}$ &  $0.22\times 10^{-11}$ &
819 GeV\\
DY-D0   &  $(0.00 \pm 0.12)\times 10^{-11}$ &  $0.15\times 10^{-11}$ &
906 GeV\\
DYcombined &  $(0.00 \pm 0.10)\times 10^{-11}$& $0.14\times 10^{-11}$ &
925  GeV\\
\hline
LEPII hadronic &  $(0.45$ \err{0.27}{1.18})$\times 10^{-11}$ & 
$0.82\times 10^{-11}$ & 591  GeV \\
LEPII leptonic &  $(-0.34$ \err{0.80}{0.68}$)\times 10^{-11}$ & 
$1.01\times 10^{-11}$ & 561  GeV \\
LEPII b,c &  $(1.63$ \err{1.69}{5.87}$)\times 10^{-11}$ & 
$4.23\times 10^{-11}$ & 392  GeV \\
LEPII   &  $(-0.44$ \err{1.01}{0.29}$)\times 10^{-11}$ & 
$0.74\times 10^{-11}$ & 607  GeV \\
\hline
HERA    &  $(-0.050$ \err{0.18}{0.17}$)\times 10^{-11}$ & 
$0.47\times 10^{-11}$       &  679 GeV\\
\hline
\hline
ALL    &   $(-0.026$ \err{0.11}{0.076}$)\times 10^{-11}$ & 
$0.13\times 10^{-11}$       &  939 GeV\\
ALL w/o HERA & $(-0.0055$ \err{0.11}{0.10}$)\times 10^{-11}$ & 
$0.14\times 10^{-11}$       &  925 GeV\\
\hline
\hline
\end{tabular}
\end{table}

\end{document}